# Unconventional superconductivity in twisted bilayer WSe$_2$

Yiyu Xia$^{1*}$, Zhongdong Han$^{2*}$, Kenji Watanabe$^{3}$, Takashi Taniguchi$^{3}$, Jie Shan$^{1,2,4**}$, Kin Fai Mak$^{1,2,4**}$

$^{1}$School of Applied and Engineering Physics, Cornell University, Ithaca, NY, USA
$^{2}$Laboratory of Atomic and Solid State Physics, Cornell University, Ithaca, NY, USA
$^{3}$National Institute for Materials Science, Tsukuba, Japan
$^{4}$Kavli Institute at Cornell for Nanoscale Science, Ithaca, NY, USA

*These authors contributed equally
\*\*Email: jie.shan@cornell.edu; kinfai.mak@cornell.edu

**Moiré materials have enabled the realization of flat electron bands and quantum phases that are driven by strong correlations associated with flat bands[1-5]. Superconductivity has been observed, but solely, in graphene moiré materials[6-11]. The absence of robust superconductivity in moiré materials beyond graphene, such as semiconductor moiré materials[5], has remained a mystery and challenged our current understanding of superconductivity in flat bands. Here, we report the observation of robust superconductivity in 3.65-degree twisted bilayer WSe$_2$ which hosts a honeycomb moiré lattice[12,13]. Superconductivity emerges at half-band filling and under small sublattice potential differences, where the moiré band is a flat Chern band[12,13]. The optimal superconducting transition temperature is about 220 mK and constitutes 2% of the effective Fermi temperature; the latter is comparable to the value in high-temperature cuprate superconductors[14,15] and suggests strong pairing. The superconductor borders on two distinct metals below and above half-band filling; it undergoes a continuous transition to a correlated insulator by tuning the sublattice potential difference. The observed superconductivity on the verge of Coulomb-induced charge localization suggests roots in strong electron correlations[14,16].**

**Main**

The discovery of superconductivity in twisted bilayer graphene[6] has initiated intense research on moiré superlattices of van der Waals materials[1,2,4,5]. In particular, transition metal dichalcogenide (TMD) semiconductors, MX$_2$ (M = Mo, W; X = S, Se, Te), in the monolayer limit can be viewed as gapped graphene with strong Ising spin-orbit coupling[17]; TMD moiré materials have emerged as a simple yet extremely rich model system for studies of correlated and topological phases of matter[5]. The tunable moiré flat bands in these materials, which strongly enhance the correlation effects, have stabilized the Mott insulators[18-20], generalized Wigner crystals[18,21] and heavy fermions[22]. The combined correlation and non-trivial band topology have further induced the integer and fractional Chern insulators[23-28] and the fractional quantum spin Hall insulator[29], the latter of which has not been observed in any other materials. However, superconductivity—a hallmark of graphene flat band systems both with and without the moiré effects[6-11,30-34]—has remained elusive in TMD moiré materials. An earlier study reported the observation of a potential superconducting state in twisted bilayer WSe$_2$ (tWSe$_2$) by doping a correlated insulator[20], but the state appears unstable to repeated thermal cycles.

In this work, we report an electrical transport study of 3.65°-tWSe$_2$. We observe robust superconductivity on the verge of Coulomb-induced charge localization around half-band filling when the interlayer potential difference is tuned close to zero. The state borders on two distinct metals, below and above half-band filling, respectively. The superconducting transition temperature constitutes about 2% of the effective Fermi temperature, which is comparable to the value in cuprate high-temperature superconductors[14,15] and suggests strong pairing. The observed superconducting state does not stem from doping a correlated insulator and cannot be readily explained by the existing theories[35-49]. Future experiments and theories are required to fully understand the nature of the state.

Figure 1a illustrates a schematic of the dual-gated tWSe$_2$ device employed in this study. Twisted bilayer WSe$_2$ has a honeycomb moiré lattice with two sublattice sites residing at the MX and XM stacking regions[12,13] (Fig. 1b). The moiré density [$n_M = (4.25 \pm 0.03) \times 10^{12}$ cm$^{-2}$], or equivalently, the twist angle (3.65° ± 0.01°) is calibrated through quantum oscillations under a high magnetic field (Extended Data Fig. 2). The top and bottom gates are made of multilayer hexagonal boron nitride (hBN) and graphite. They independently tune the hole moiré filling factor $v$ and the electric field $E$ perpendicular to tWSe$_2$ (or equivalently, the interlayer/sublattice potential difference). The narrower top gate defines the device channel and additional Pd split gates are patterned to turn off any parallel conduction channels. To achieve ohmic contacts to tWSe$_2$ down to mK temperatures, we use Pt contact electrodes and Pd contact gates. The contact gates induce heavy hole doping in the tWSe$_2$ regions immediately adjacent to the Pt electrodes. Contact resistance between 10-40 kΩ has been achieved (Extended Data Fig. 1). See Methods for details on the device fabrication, twist angle and disorder calibration and electrical measurements.

Figure 1c illustrates the electronic band structure for the first two moiré valence bands of 3.65°-tWSe$_2$ at $E = 0$, calculated using a reported continuum model[13] (Methods). The bands are composed of the spin-valley locked electronic state from the K or K' valley of monolayer WSe$_2$. Only the K-valley state is shown in Fig. 1c. Both moiré bands carry Chern number +1 (-1 for the K'-valley state), and are expected to transition to non-topological bands under a sufficiently high electric field that overcomes the interlayer hopping[12,13]. Figure 1d displays the corresponding electronic density of states (DOS) at the Fermi level as a function of $E$ and $v$. For $E = 0$, the DOS shows a van Hove singularity (vHS) near $v = 0.75$ which arises from a saddle point located at the $m$-point of the moiré Brillouin zone[20]. Across the vHS, the hole Fermi surface changes abruptly from disconnected κ/κ' hole pockets to a single γ electron pocket (insets). As $E$ increases, the vHS shifts continuously towards higher $v$ (Extended Data Fig. 4) and exhibits a wing-like feature in the DOS.

Figure 1e shows the longitudinal resistance $R$ measured as a function of $E$ and $v$ (< 2) at 10 K. The map is dominated by a large resistance region that is centered around $E = 0$ and expands with increasing $v$. Multiple correlated insulating states ($v = 1/4$, 1/3 and 1) can be identified. In line with earlier studies[12,13,24-27], this is the topological layer-hybridized region. The region at higher electric fields is characterized by a metallic state and is the non-topological layer-polarized region. Upon further cooling of the sample down to 50 mK (Fig. 1f), all the features become sharper, allowing us to draw the dashed lines to guide the eye for the phase boundary. The resistance maps in general show good agreement with the DOS map in Fig. 1d except at commensurate fillings ($\nu = 1/4$, 1/3 and 1), where the correlation effects dominate. The enhanced resistance near $\nu = 0.75$ and $E = 0$ and at the wing-like features for $\nu > 1$ near the phase boundary follows closely the location of the calculated vHS (the enhanced resistance is presumably from the large DOS and/or enhanced scattering rate near the vHS). The assignment is further supported by a sign change of the Hall resistance (Extended Data Fig. 5). However, the most striking feature of Fig. 1f is the opening of a short strip with nearly vanishing resistance in the middle of the correlated insulator state at half-band filling ($\nu = 1$).

**Superconductivity at half-band filling**
We zoom in on the phase space near $\nu = 1$ and $E = 0$ in Fig. 2. The dashed lines are provided to guide the eye for the strip with substantially suppressed resistance at 50 mK. Zero resistance is observed near the two ends of the strip (Fig. 2a). The zero-resistance state is independent of the measurement configuration for a homogenous moiré area of 1.5 μm × 8 μm (Extended Data Fig. 1 and 3). It is robust against repeated thermal cycles; a total of 8 thermal cycles was involved in collecting the data presented in the main text. On the other hand, the zero-resistance state is susceptible to both thermal excitations and magnetic field threaded through the sample. For instance, the state is quenched when the sample is warmed up to 300 mK (Fig. 2b) and when an out-of-plane magnetic field of 50 mT is applied at 50 mK (Fig. 2c). Figure 2 also shows that the zero-resistance state is not connected to the vHS near $\nu = 0.75$ and $E = 0$. It borders on two distinct metallic states, more discussions on which will follow in the next section.

For now, we focus on the zero-resistance state at the upper end of the strip ($\nu \approx 1$ and $E \approx 8$ mV/nm) and examine its response to bias current $I$, temperature $T$ and magnetic field $B$. Figure 3a displays the differential resistance, $\frac{dV}{dI}$, as a function of $T$ and $I$ in the absence of magnetic field. Linecuts at representative temperatures are illustrated in Fig. 3b. Below about 180 mK (dashed line in Fig. 3b), bias current above a critical value is required to destroy the zero-resistance state. The critical current is about 5 nA at 50 mK, and the value decreases monotonically with increasing temperature. Above about 250 mK, the differential resistance increases slightly with bias. Between the two temperatures, a resistance dip at zero bias is still observed although zero resistance is no longer reached. Figure 3c shows the temperature dependence of the differential resistance at zero bias, $R$. As temperature decreases, $R$ drops to zero in two steps, including an initial rapid descend followed by the onset of zero resistance.

A similar study on the magnetic-field response is shown in Fig. 3d-f. Figure 3d displays the differential resistance, $\frac{dV}{dI}$, as a function of $B$ and $I$ at 50 mK (linecuts at representative magnetic fields are included in Extended Data Fig. 6d). Similar to Fig. 3a,b, the critical current required to destroy the zero-resistance state vanishes continuously with increasing magnetic field. The zero-resistance critical field, $B_{C1}$, is about 6 mT. Above this field, a resistance dip at zero bias is still observed and disappears in the normal state above the second critical field $B_{C2}$. Figure 3e shows the zero-bias resistance, $R$, as a function of $T$ and $B$. Linecuts at representative magnetic fields are shown in Fig. 3f. At 50 mK, the normal state is reached at about 80 mT, above which $R$ has a weak field dependence. Between the two critical fields, $R$ increases linearly with field, and $B_{C2}$ is defined as the field at which the linear fits to the field dependence of $R$ in two different phases intersect (Extended Data Fig. 6c). The critical fields are evaluated for all temperatures (filled symbols, Fig. 3e). They vanish continuously at the corresponding transition temperatures.

The results above are fully consistent with superconductivity in two dimensions (2D). Because of the enhanced thermal fluctuations, the superconducting transition in 2D is expected to occur over a broad temperature range in two steps[50] as observed in experiment. The two steps are understood as the onset of Cooper pairing at the pairing temperature, $T_P$, and the onset of quasi-long-range phase coherence at the Berezinskii-Kosterlitz-Thouless (BKT) transition temperature, $T_{BKT}$. We define $T_P$ ($\approx$ 250 mK) as the temperature at which the zero-bias resistance reaches 80% of the normal-state resistance (Extended Data Fig. 6a). We determine $T_{BKT}$ ($\approx$ 180 mK) by performing the BKT analysis (inset of Fig. 3c), that is, the nonlinear $I - V$ dependence takes the form $V \propto I^3$ (dashed line) at $T_{BKT}$ (Ref. [50,51]).

The observed broad superconducting transition induced by magnetic fields is also expected for weakly pinned vortices. Here $B_{C1}$ and $B_{C2}$ correspond to the critical fields, at which the zero-resistance state and Cooper pairing with finite resistance are destroyed, respectively. The observed $R \propto \frac{B}{B_{C2}}$ between the two critical fields is consistent with nearly unpinned vortices in the superconductor[50]. Such mobile vortices have been reported in 2D crystalline superconductors with shallow pinning sites[51]. Our high-quality tWSe$_2$ device with low disorder density (about 2.5% of the moiré density $n_M$, Extended Data Fig. 1) is compatible with this picture. We can estimate the superconducting coherence length $\xi$ by fitting the Ginzburg-Landau result $B_{C2} \approx \frac{\Phi_0}{2\pi \xi^2}\left(1 - \frac{T}{T_P}\right)$ to the experimental temperature dependence of $B_{C2}$ near $T_P$ (dashed line in Fig. 3e, Ref. [50]). Here $\Phi_0 = h/2e$ is the flux quantum with $h$ and $e$ denoting the Planck constant and fundamental charge, respectively. The superconducting coherence length ($\xi \approx$ 52 nm) is about 10 times of the moiré period ($a_M \approx$ 5 nm).

**Doping dependence of the superconductor**
We examine the doping dependence of the superconducting state, focusing on the upper end of the strip (Fig. 2a, $E \approx$ 8 mV/nm) as above. Figure 4a shows zero-bias resistance, $R$, as a function of $T$ and $\nu$ in the absence of magnetic fields. Linecuts at representative temperatures are shown in Fig. 4b. Superconductivity is observed only in the vicinity of $\nu = 1$ with no more than 0.04 filling on each side. The pairing and BKT transition temperatures are shown as blue and orange symbols in Fig. 4a, respectively. The optimal $T_{BKT}$ is slightly below 200 mK at $\nu \approx 1$.

The state on both sides of the superconductor is metallic. Figure 4c illustrates the temperature dependence of $R$ at three representative filling factors, $\nu = 0.9$, 1 and 1.1.

Figure 4d displays $R$ as a function of $T^2$. The metallic state above filling one shows a Fermi liquid behavior with $R = R_0 + AT^2$ (dashed line, Fig. 4d) over an extended temperature range in the low-temperature limit. Here $R_0$ is the residual resistance and $A$ is the Kadowaki-Woods coefficient that reflects the quasiparticle effective mass[52]. We define coherence temperature, $T_{coh}$, as the temperature at which $R$ deviates from the Fermi liquid behavior by 10%. As filling factor approaches one from above, $T_{coh}$ rapidly decreases (Fig. 4a), and correspondingly, $A$ or the quasiparticle effective mass rapidly increases (inset of Fig. 4d).

On the other hand, the metallic state below filling one and the normal state of the superconductor are not compatible with a Fermi liquid for nearly the entire temperature window and $T_{coh}$ cannot be reliably extracted. The normal state of the superconductor at $\nu = 1$ also exhibits a Pomeranchuk effect[53], that is, there is a resistance peak at elevated temperature $T^*$ (Fig. 4b,c). This temperature scale separates the coherent and incoherent transport regimes (below and above $T^*$, respectively) and is used as a measure of the Fermi temperature, $T_F$ (Ref. [54]). We illustrate $T^*$ as yellow symbols in Fig. 4a for all filling factors that can be identified within the measurement window of 25 K. It reaches a minimum of about 10 K at $\nu = 1$. The result shows that the normal state of the superconductor is developed from a poor metal with the lowest $T_F$, one that is on the verge of charge localization induced by the strong electronic correlation at $\nu = 1$.

**Superconductor-insulator transition**
The proximity to charge localization of the normal state of the superconductor is further supported by the superconductor-insulator transition induced by out-of-plane electric field $E$ at $\nu = 1$. Figure 5a shows zero-bias resistance, $R$, as a function of $T$ and $E$ with linecuts at representative electric fields shown in Fig. 5b. All resistance curves merge into the same temperature dependence above about 4 K (see Fig. 1e for a wider $E$-field range at 10 K). A sharp superconductor-insulator transition is observed near the critical electric field $E_C \approx 11.7$ mV/nm. The corresponding resistance $R_C$ (dashed line in Fig. 5b) has the weakest temperature dependence.

Figure 5c demonstrates the collapse of the resistance curves in the vicinity of $E_C$. In this process, we first extract the thermal activation gap $T_0$ of the insulating state at each $E$ (Extended Data Fig. 7). The activation gap (black symbols, Fig. 5a) vanishes continuously as $E$ approaches $E_C$ from above. The normalized resistance $R/R_C$ for all fields collapses into two groups of curves if we scale the temperature axis by $T_0$. For the superconducting state, we have used the same $T_0$ as its insulating counterpart that lies with equal distance to $E_C$. One group of the collapsed curves decreases, and the other group diverges, with decreasing $T/T_0$. In Fig. 5a, we also show the electric-field dependence of $T_{BKT}$ and $T_P$ on the superconducting side. Both temperature scales vanish continuously as $E$ approaches $E_C$ from below. The continuously vanishing temperature/energy scales and the collapse of the resistance curves in the vicinity of $E_C$ suggest that the superconductor-insulator transition is a continuous quantum phase transition tuned by the out-of-plane electric field or interlayer/sublattice potential difference.

**Concluding remarks**

We have observed robust superconductivity in twisted bilayer WSe$_2$, solving a long-standing mystery in the research of moiré semiconductors. The observed superconducting state has several unusual properties that deserve future studies. First, superconductivity is observed only in the topological layer-hybridized region of the twisted bilayer[12,13,24-27] (Fig. 1f). The TMD moiré system with tunable band topology is an excellent platform to investigate the role of band topology and quantum geometry in superconductivity[55]. Second, superconductivity is strongly confined near $\nu = 1$ and is away from the vHS (at $\nu \approx 0.75$ near $E = 0$). The superconductor evolves continuously to a correlated insulator by tuning the interlayer potential difference (Fig 2 and 5). The phenomenology is different from that in graphene moiré systems, where superconductivity emerges often by doping a correlated insulator[6-11]. Third, the normal state of the superconductor is a strongly correlated metal with minimal $T_F \approx T^*$ on the verge of localization (Fig. 4). The superconducting state borders on two distinct metallic states: a well-behaved Fermi liquid above filling one and one that largely deviates from the Fermi-liquid behavior below filling one. These metallic states are not spontaneously spin/valley-polarized (Extended Data Fig. 5), also distinct from the phenomenology of superconductivity in bilayer and trilayer graphene[31,33,34,56]. Finally, the observed superconductor is in the strong pairing limit with $T_C$ (superconducting transition temperature) about 2% of $T_F$ and superconducting coherence length about 10 times of the moiré period (see Methods for additional estimates). These values are comparable to that in high-$T_C$ cuprates[14,57]. Transport experiments on samples with varying twist angles and different TMDs, as well as magnetic circular dichroism measurements, could help to address questions such as the nature of the correlated insulator[58,59] and the normal metallic states surrounding the superconductor. Our experiment opens the door to explore unconventional superconductivity driven by strong electronic correlations in semiconductors moiré materials.

## Methods
### Device fabrication.
Extended Data Fig. 1a shows a schematic representation of dual-gated tWSe$_2$ devices. They were fabricated using the 'tear-and stack' and layer-by-layer dry transfer method, as described in previous studies[6,60]. In short, flakes of few-layer graphite, multilayer hexagonal Boron Nitride (hBN) and monolayer WSe$_2$ were first exfoliated from bulk crystals onto Si/SiO$_2$ substrates and identified based on their optical reflection contrast. The WSe$_2$ flake was cut into two halves using an AFM (atomic force microscope) tip. The flakes were then picked up sequentially by a polycarbonate thin film on a PDMS (polydimethylsiloxane) stamp in the following order: hBN, top-gate graphite, hBN, one half of the WSe$_2$ flake, second half of the WSe$_2$ flake with a twist, hBN and bottom-gate graphite. The complete stack was subsequently released onto a Si/SiO$_2$ substrate with pre-patterned Pt electrodes in the Hall bar geometry at 200°C. Finally, the contact gates and the split gates were added using standard electron-beam lithography and evaporation of Ti/Pd (5 nm/40 nm in thickness). The top gate is smaller than the bottom gate in area and defines the device channel. The contact gates serve to heavily hole-dope the tWSe$_2$ regions between the channel and the Pt electrodes to reduce the contact resistance (even when the doping density in the channel is low). The split gates are used to deplete the parallel tWSe$_2$ regions that are gated only by the bottom gate. An optical micrograph of the device presented in the main text is shown in Extended Data Fig. 1b.

**Electrical measurements.**
The electrical transport measurements were performed in a dilution refrigerator (Bluefors LD250) equipped with a 12 T superconducting magnet. Low-temperature RC and RF filters (QDevil) were installed on the mixing chamber plate to filter out the electrical noise from about 65 kHz to tens of GHz. A 1 MΩ resistor was added in series to limit the excitation current. Voltage pre-amplifiers with large input impedance (100 MΩ) were used to measure sample resistances up to about 10 MΩ. Low-frequency (5.777 Hz) lock-in techniques were adapted to measure the sample resistance with a small excitation current (< 10 nA) to avoid sample heating. Specifically, the excitation current was fixed below 1 nA to probe the superconducting state. Both the voltage drop at the probe electrodes and the source-drain current were recorded. All data were taken at the base temperature (~ 50 mK) unless specified otherwise.

**Twist angle calibration.**
The twist angle of WSe$_2$ was calibrated through the quantum oscillations observed under a perpendicular magnetic field of 12 Tesla. We first acquired a resistance map under zero magnetic field as a function of the top gate and bottom gate voltages ($V_{tg}$ and $V_{bg}$, respectively). We converted the gate voltages to a perpendicular electrical field, $E = \frac{1}{2}\left(\frac{V_{bg}}{d_{bg}} - \frac{V_{tg}}{d_{tg}}\right)$, and filling factor, $\nu \propto \frac{V_{bg}}{d_{bg}} + \frac{V_{tg}}{d_{tg}}$, using the hBN thickness in the top and bottom gates ($d_{tg}$ and $d_{bg}$, respectively). The hBN thicknesses were independently calibrated by AFM, and their ratio was fine-tuned to align the resistance peaks parallel to the electric-field axis (Extended Data Fig. 2a). We assigned the filling factor of the two most prominent resistance peaks to be $\nu = 1$ and $\nu = 2$. Using the same thickness values, we also obtained the resistance map at 12 T as a function of $E$ and $\nu$ (Extended Data Fig. 2b). Landau levels are observed. We focus on the layer-polarized region under large electric fields, where the Landau levels are spin- and valley-polarized akin to that in hole-doped monolayer WSe$_2$ under large magnetic fields[61]. Landau levels with index $\nu_{LL} = 2 - 8$ are denoted by vertical dashed lines, and their filling dependence is shown in Extended Data Fig. 2c. We determined the moiré density, $n_M = (4.25 \pm 0.03) \times 10^{12}$ cm$^{-2}$, from the slope ($\frac{\phi_0}{B}n_M$ with $\phi_0 = \frac{h}{e}$ denoting the magnetic flux quantum) of the best linear fit to the experimental data. We obtained the twist angle, $\theta = a\sqrt{\frac{\sqrt{3}}{2}n_M} = 3.65° \pm 0.01°$ from the moiré density, where $a \approx 3.317$ Å (Ref. [62]) is the lattice constant of monolayer WSe$_2$.

**Band structure calculation.**
To capture the low-energy electronic band structure of tWSe$_2$ of small twist angle, we study the continuum model based on the effective mass description first introduced by Wu *et al.* (Ref. [12]). In monolayer TMDs, the topmost valence bands are spin split by 100's meV and the spin and valley are locked due to inversion symmetry breaking and strong spin-orbit coupling (Ref. [17]). This property is inherited by tWSe$_2$. For each valley, the low-energy physics of tWSe$_2$ can be described by a two-band $k \cdot p$ model with a periodic pseudomagnetic field, $\Delta(\mathbf{r}) = (\text{Re}\Delta_T^\dagger, \text{Im}\Delta_T^\dagger, \frac{\Delta_b - \Delta_t}{2})$, where $\Delta_T$ is the interlayer tunneling amplitude and $\Delta_{b,t}$ are the bottom and top layer dependent energies. The effective moiré Hamiltonian for the $+K$ valley state (with spin-↑) is given by

$$H_\uparrow = \begin{pmatrix} -\frac{\hbar^2 (k-\kappa_+)^2}{2m^*} + \Delta_b(\mathbf{r}) & \Delta_T(\mathbf{r}) \\ \Delta_T^\dagger(\mathbf{r}) & -\frac{\hbar^2 (k-\kappa_-)^2}{2m^*} + \Delta_t(\mathbf{r}) \end{pmatrix}, \quad (1)$$

where $m^*$ is the valence band effective mass (= $0.45m_0$ for monolayer WSe$_2$ [Ref. 63]) and $\kappa_\pm$ are the corners of the moiré Brillouin zone (mBZ). The effective Hamiltonian for the $-K$ valley state (with spin-↓) is the complex conjugate of $H_\uparrow$.

The pseudomagnetic field with the lattice symmetry constraints can be described in the lowest harmonic approximation as

$$\Delta_{b,t}(\mathbf{r}) = \pm \frac{V_z}{2} + 2V \sum_{j=1,3,5} \cos(\mathbf{g}_j \cdot \mathbf{r} \pm \psi), \quad (2)$$

$$\Delta_T(\mathbf{r}) = w(1 + e^{-i\mathbf{g}_2 \cdot \mathbf{r}} + e^{-i\mathbf{g}_3 \cdot \mathbf{r}}). \quad (3)$$

Here $V_z$ denotes the sublattice potential difference and $\mathbf{g}_j$ is the reciprocal lattice vector obtained by counterclockwise rotations of $\mathbf{g}_1 = (\frac{4\pi}{\sqrt{3}a_M}, 0)$ by angle $(j-1)\pi/3$. The parameters $(V, \psi, w) = (9.0 \text{ meV}, 128°, 18 \text{ meV})$ are taken from the DFT (density functional theory) calculations for tWSe$_2$ (Ref. 13).

We obtain the band structure and density of states (DOS) by diagonalizing the Hamiltonian given in Eq. (1). The result for 3.65°-twisted WSe$_2$ (as studied in the experiment) is shown in Fig. 1c,d. To compare with experiment, we convert the sublattice potential difference to vertical electric field using $E = V_z / (\frac{\varepsilon_{hBN}}{\varepsilon_{TMD}} et)$. Here the dipole moment $\frac{\varepsilon_{hBN}}{\varepsilon_{TMD}} et \approx 0.26\ e \cdot nm$ is independently determined from the anti-crossing feature of the layer-hybridized moiré excitons[53,64] ($\varepsilon_{hBN} \approx 3$ and $\varepsilon_{TMD} \approx 8$ are the out-of-plane dielectric constants of hBN and TMD, respectively, and $t \approx 0.7\ nm$ is the interlayer separation between the WSe$_2$ monolayers).

**Estimate of superconductor properties.**
We estimate the ratio $\frac{T_C}{T_F}$ for superconducting tWSe$_2$ at $\nu > 1$, where $T_C$ and $T_F$ are the superconducting transition temperature and the Fermi temperature, respectively. We use the mean of the pairing and BKT transition temperatures to represent $T_C \approx \frac{T_{BKT} + T_P}{2} \approx 220$ mK and use $T^* \approx 10$ K to approximate $T_F$. The estimated value $\frac{T_C}{T_F} \approx 0.02$ suggests strong pairing and is comparable to that in high-$T_C$ cuprates[14]. The strong pairing is further evidenced by the ratio $\frac{\xi}{a_M} \approx 10$, where $a_M \approx 5$ nm is the moiré period and $\xi \approx 52$ nm is the superconducting coherence length extracted from the critical magnetic-field measurement (Fig. 3e). The equivalent value for cuprates is about 5 (Ref. [57]).

We can further estimate Fermi velocity $v_F \approx \frac{\xi k_B T_C}{\hbar} \approx 1,500$ m/s and effective quasiparticle mass $m \approx \frac{\hbar^2 \pi n_M}{k_B T_F} \approx 10\ m_0$ at the Fermi level, where $k_B$, $\hbar$ and $m_0$ denote the Boltzmann constant, reduced Planck's constant and the free electron mass, respectively. The obtained values are self-consistent with the relation $m v_F = \hbar \sqrt{2\pi n_M}$. Both $m$ and $v_F$ are strongly renormalized from their single-particle values based on the continuum model[12,13] (Fig. 1c).

**Acknowledgements**
We thank B. A. Bernevig, D. Chowdhury, S. Das Sarma, L. Fu, C. Jian, E.-A. Kim, K. T. Law, A. H. MacDonald and Q. Si for helpful discussions. We also thank J. Zhu and P. Knüppel for many technical discussions.

Figures

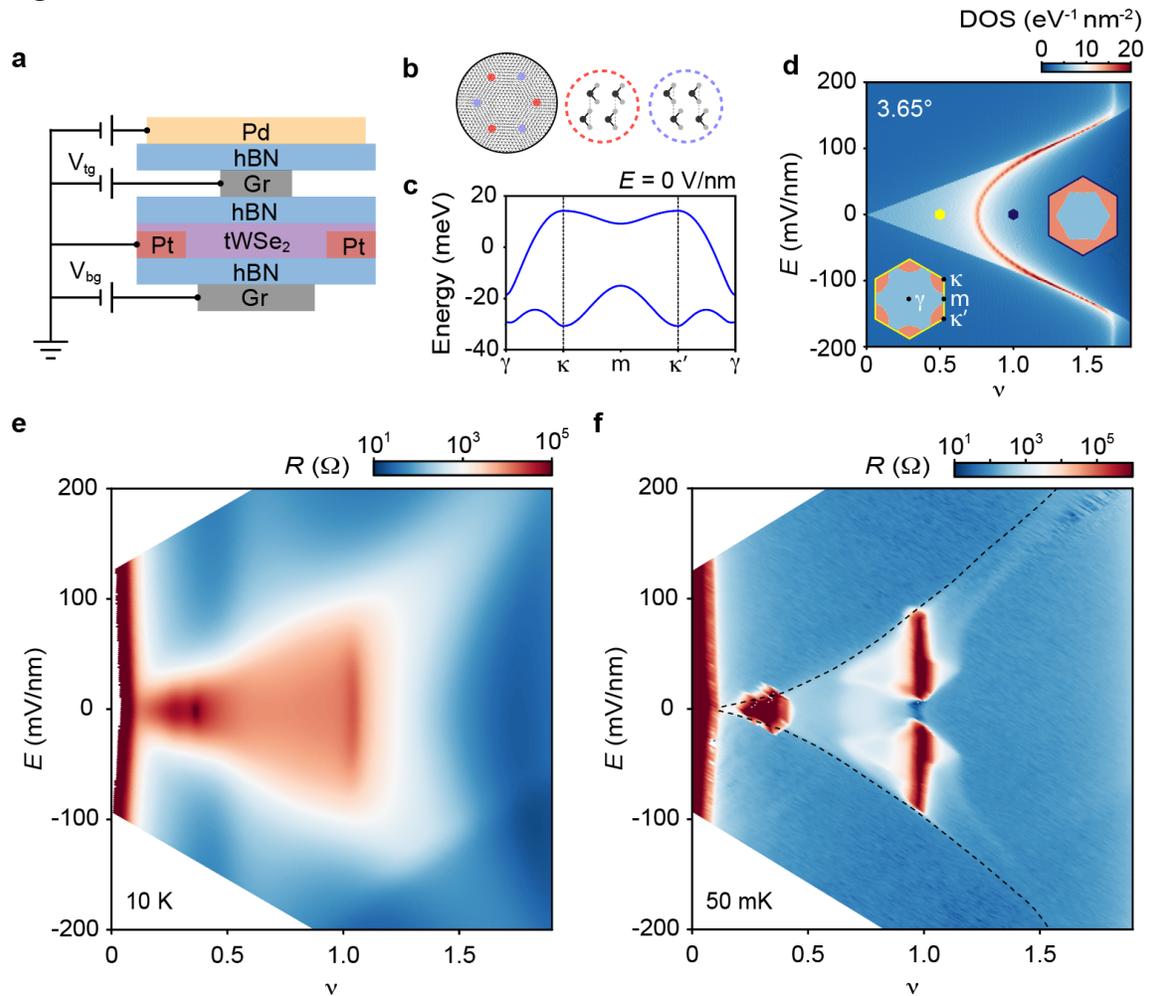

**Figure 1 | Electronic structure of tWSe$_2$. a,** Schematic of a dual-gated tWSe$_2$ device. Both gates are made of hBN and few-layer graphite (Gr) with the narrower top gate defining the tWSe$_2$ channel. The top and bottom gate voltages ($V_{tg}$ and $V_{bg}$, respectively) control the vertical electric field $E$ and hole filling factor $\nu$ in tWSe$_2$. Platinum (Pt) is the contact electrodes to tWSe$_2$. Additional palladium (Pd) contact gate and split gate voltages ($V_{contact/split}$) turn on the Pt contacts and turn off the parallel channels, respectively (only one gate is shown). **b,** Honeycomb moiré lattice of tWSe$_2$ with sublattice sites centered at the MX (red) and XM (blue) stacking sites; black and grey dots denote the M (= W) and X (= Se) atoms, respectively. **c,** Topmost moiré valence bands for the K-valley state of 3.65°-tWSe$_2$ from the continuum model. Both bands carry Chern number +1. The corresponding bands of the K'-valley state carry Chern number -1. **d,** Electronic density of states (DOS) versus $E$ and $\nu$. The vHS ($\nu \approx 0.75$ at $E = 0$) disperses towards higher $\nu$ with increasing $E$. Insets: Fermi surface at $E = 0$ evolves from disconnected hole pockets centered at κ/κ' of the moiré Brillouin zone (yellow) to a single electron pocket centered at γ (black) as $\nu$ passes the vHS. **e,f,** Longitudinal resistance $R$ as a function of $E$ and $\nu$ at 10 K (**e**) and 50 mK (**f**). The dashed lines in **f** (a guide to the eye) separate the layer-hybridized and layer-polarized regions.

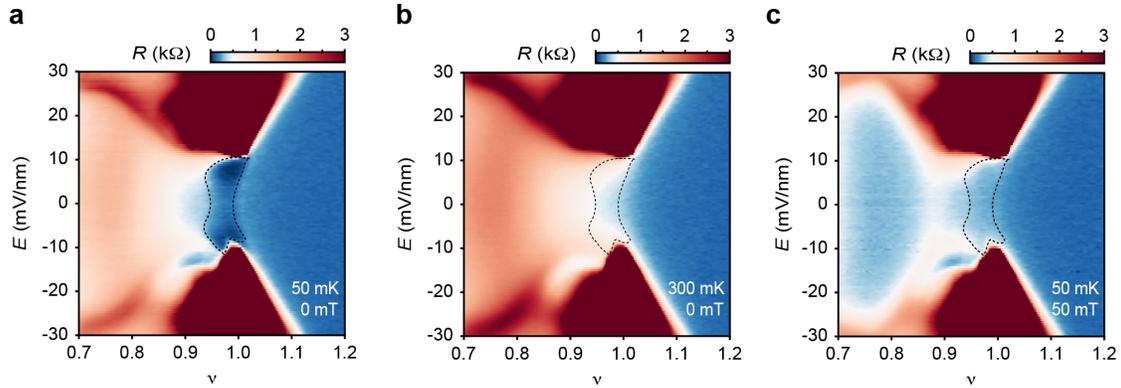

**Figure 2 | Zero-resistance region around half-band filling. a-c,** Longitudinal resistance $R$ as a function of $E$ and $\nu$ near $E = 0$ and $\nu = 1$ at different temperatures and externally applied magnetic fields. The dotted lines are a guide to the eye of the zero-resistance region observed at 50 mK (**a**). The zero-resistance state is quenched by increasing the sample temperature to 300 mK (**b**) or by applying an out-of-plane magnetic field of 50 mT (**c**). The metallic states immediately below and above half-band filling are insensitive to the magnetic field in **c**.

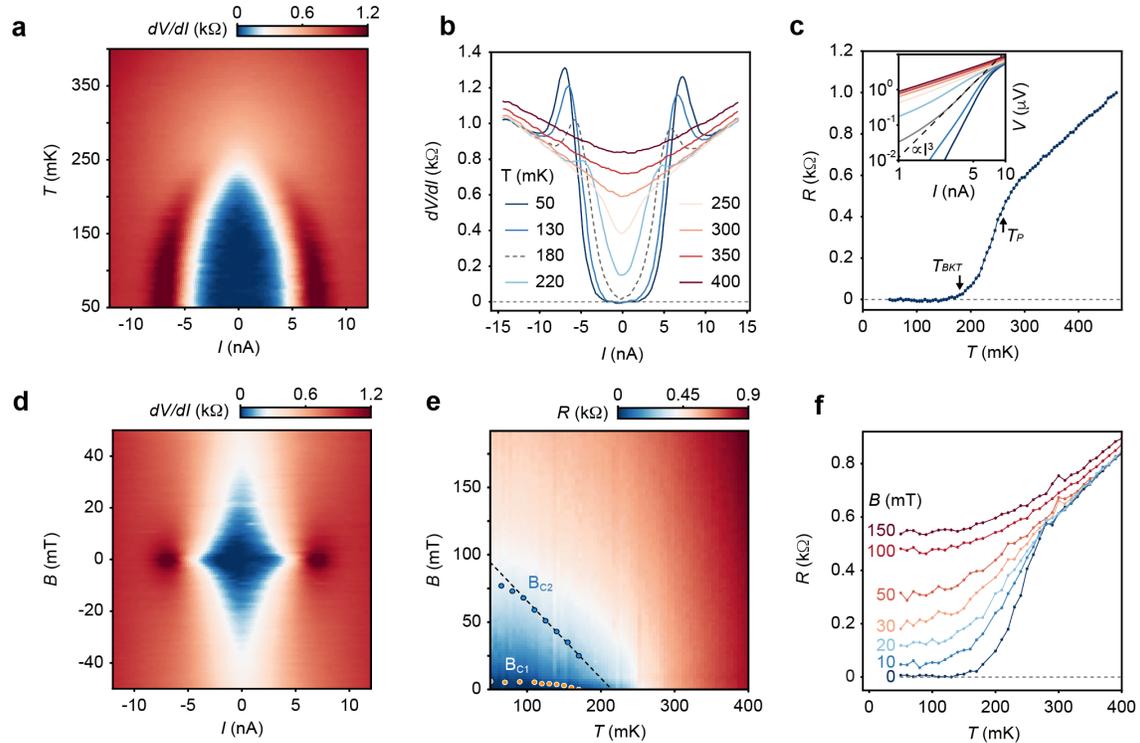

**Figure 3 | Superconductivity at half-band filling. a,** Differential resistance, $\frac{dV}{dI}$, as a function of temperature $T$ and bias $I$ under zero applied magnetic field. **b,** Linecuts of **a** at representative temperatures. The critical current vanishes continuously with increasing temperature. Dashed line: differential resistance at the BKT transition ($T_{BKT} \approx 180$ mK). **c,** Temperature dependence of zero-bias resistance $R$ with two temperature scales, $T_{BKT}$ and $T_P$. Inset: at $T_{BKT}$, the $V$-$I$ dependence follows $V \propto I^3$ (dashed line); the line color is defined in **b**. At $T_P (\approx 250$ mK), $R$ reaches about 80% of the normal-state resistance. **d,** Differential resistance as a function of $B$ and $I$ at 50 mK. The critical current vanishes continuously with increasing magnetic field. **e,** Zero-bias resistance $R$ as a function of $B$ and $T$ with critical field $B_{C1}$ and $B_{C2}$. The dashed line is a fit of $B_{C2} \approx \frac{B}{2\pi\xi^2}\left(1 - \frac{T}{T_P}\right)$ to data, from which the superconducting coherence length $\xi \approx 52$ nm is determined. **f,** Linecuts of **e** at representative magnetic fields. All data in Fig. 3 are obtained for $\nu \approx 1$ and $E \approx 8$ mV/nm.

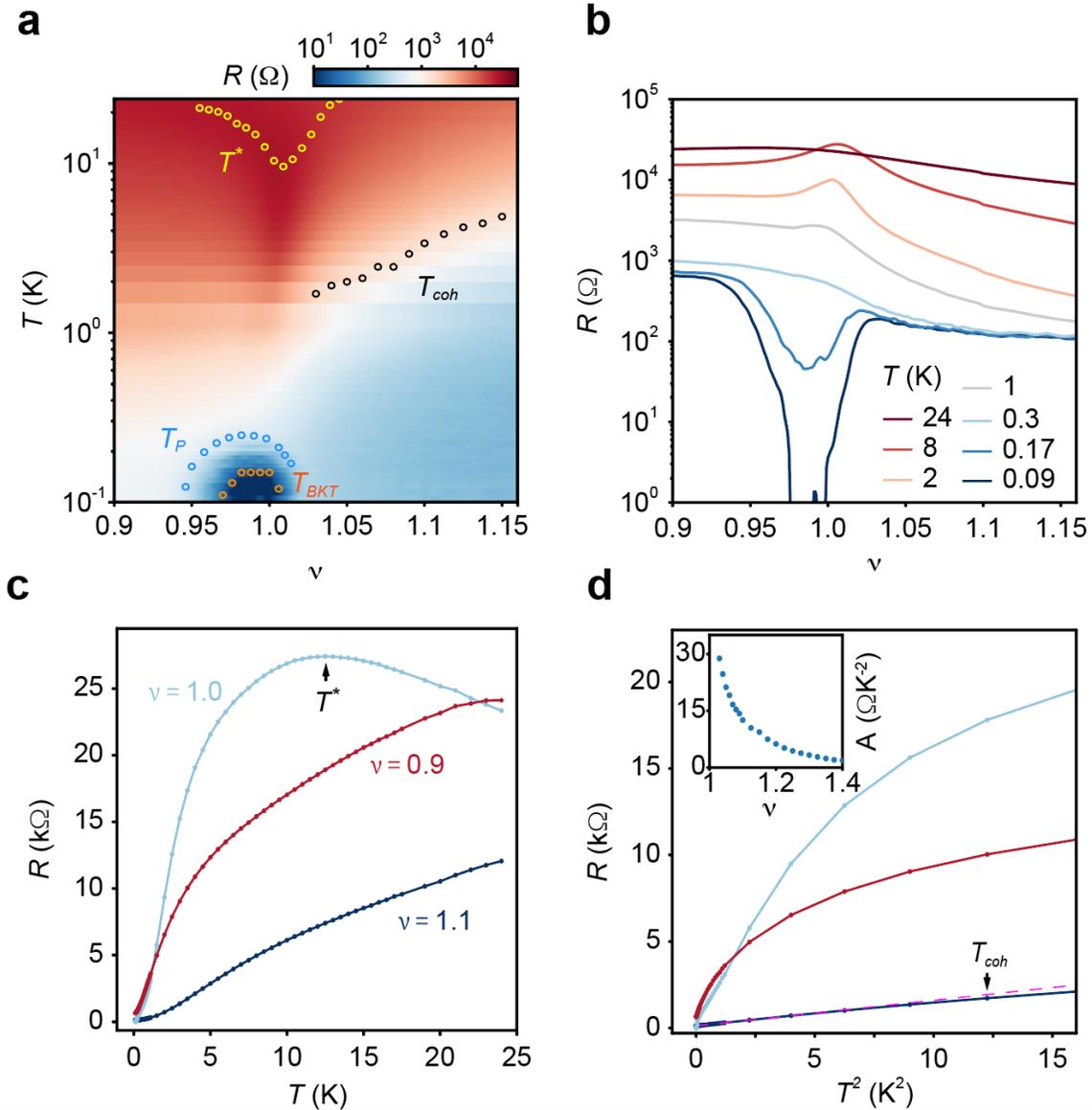

**Figure 4 | Doping dependence of the superconducting state. a,** Zero-bias resistance $R$ as a function of $T$ and $\nu$ at $E \approx 8$ mV/nm and $B = 0$. Superconductivity is observed only near $\nu = 1$. The corresponding normal state shows a resistance peak around 10 K. **b,c,** Linecuts of **a** at representative temperatures (**b**) and filling factors (0.9, 1.0 and 1.1) (**c**). In **c**, $T^*$ denotes the temperature corresponding to the resistance maximum. **d,** Same as **c** up to 4 K displayed as a function of $T^2$. The dependence for $\nu > 1$ is described by $R = R_0 + AT^2$ (dashed line), where residual resistance $R_0$ and coefficient $A$ are free parameters. At $T_{coh}$, $R$ deviates from the $T^2$-dependence by 10%. Inset: filling dependence of $A$.

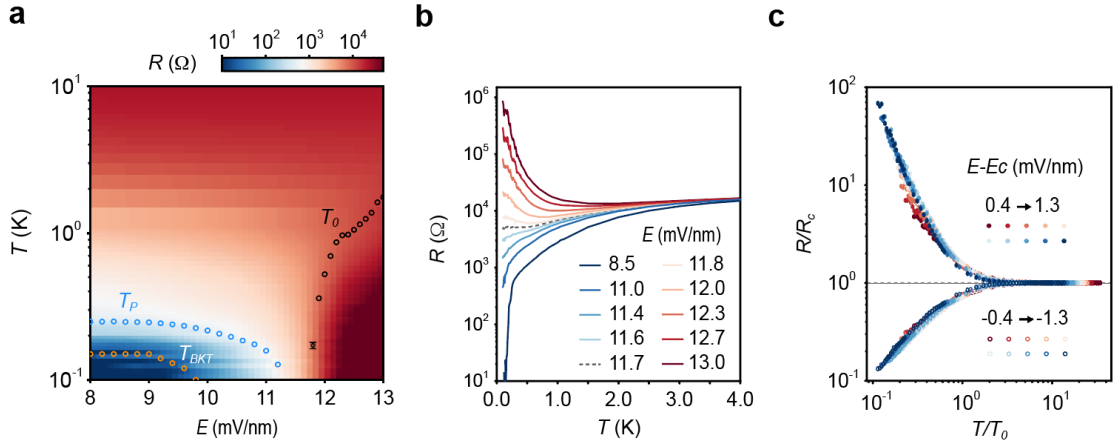

**Figure 5 | Superconductor-to-insulator transition. a,** Zero-bias resistance $R$ as a function of $T$ and $E$ at $\nu \approx 1$ under zero magnetic field. **b,** Linecuts of **a** at representative electric fields. A superconductor-insulator transition is observed near critical field $E_C \approx 11.7$ mV/nm, at which resistance $R_C$ has the weakest temperature dependence (dashed line). **c,** Collapse of normalized resistance $R/R_C$ into two groups after scaling $T$ by $T_0$. Here $T_0$ is the thermal activation temperature extracted from experiment for each $E > E_C$; the same value of $T_0$ is used for scaling for $E < E_C$ with the same distance to $E_C$. The colors denote different electric fields measured from $E_C$ with a step size of 0.1 mV/nm; the filled and empty symbols denote $E$ above and below $E_C$, respectively.

**Extended Data Figures**

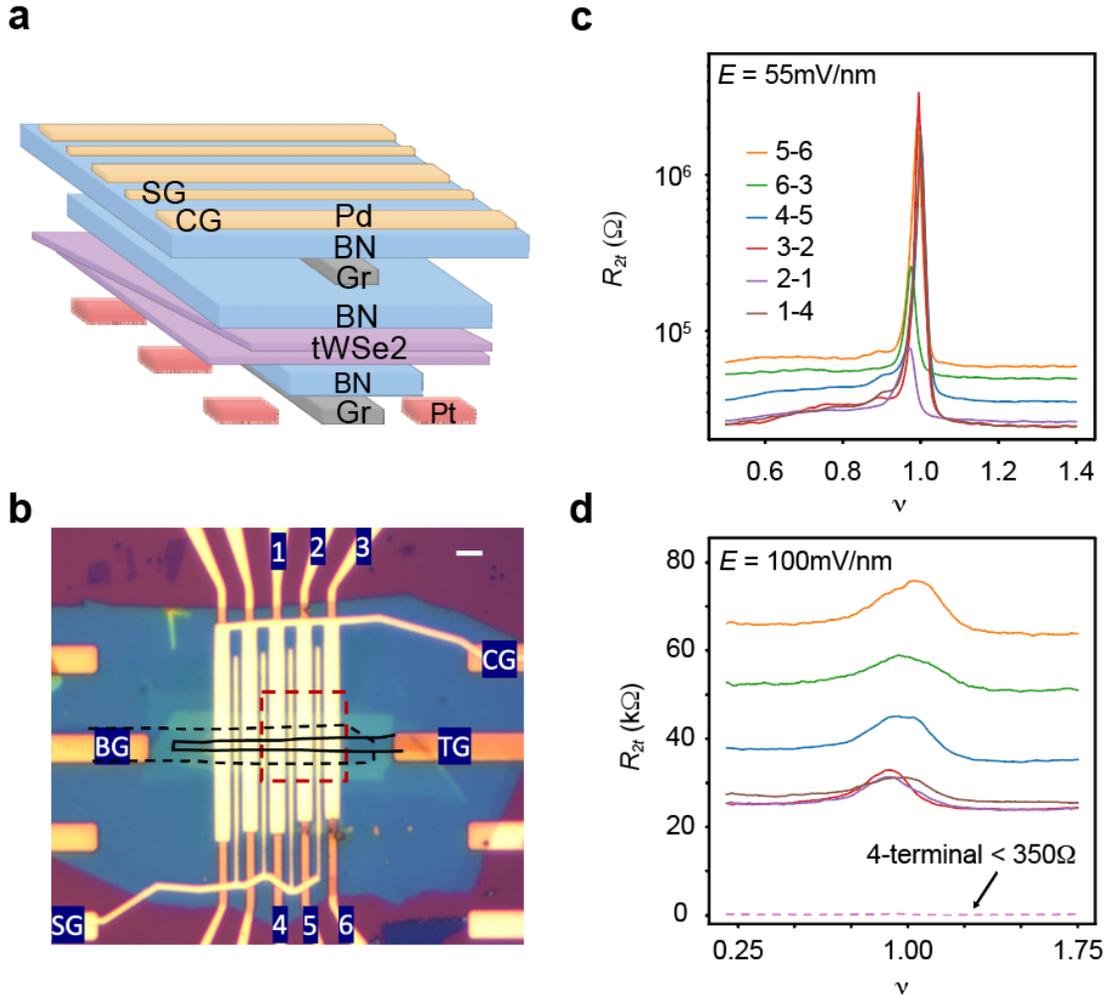

**Extended Data Figure 1 | Sample and device characterization. a,** Three-dimensional schematic of a tWSe$_2$ dual-gated device. Both the top gate (TG) and bottom gate (BG) are made of hBN and few-layer graphite (Gr). The tWSe$_2$ sample is contacted by Pt electrodes. The Pd contact gates (CG) and split gates (SG) turn on the Pt contacts and turn off the parallel channels, respectively. **b,** Optical micrograph of the device used in this study. Specific features of interest are BG (enclosed by the black dashed line), TG (black solid line), uniform moiré region (red dashed line) and Pt contact electrode 1-6. The scale bar is 4 μm. **c,d,** Filling factor dependence of two-terminal resistance $R_{2t}$ for different contact pairs at $T = 1$ K and B = 0 T. The sample is an insulator at $\nu \approx 1$ for $E$ = 55 mV/nm (**c**). The variation in filling factor for the resistance peak is about 0.025, which corresponds to a disorder density of $1 \times 10^{11}$ cm$^{-2}$. The sample is a metal at $\nu \approx 1$ for $E$ = 100 mV/nm (**d**) and the four-terminal resistance is below 350 Ω. The contact resistance is determined from $R_{2t}$ to be 10-40 kΩ.

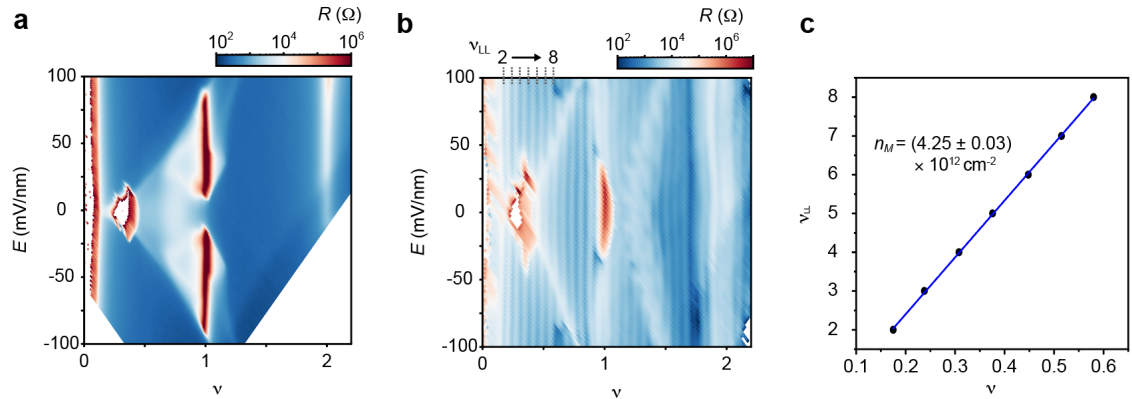

**Extended Data Figure 2 | Calibration of the moiré density. a,b,** Longitudinal resistance $R$ as a function of $E$ and $\nu$ at 50 mK under $B = 0$ T (**a**) and 12 T (**b**). Large bias current is used in the measurement and superconductivity is not observed in **a**. Landau levels are clearly observed in **b**. Landau levels with index $\nu_{LL} = 2 - 8$ (denoted by dotted lines) in the layer-polarized region are spin- and valley-polarized (i.e. nondegenerate). **c,** Landau level index $\nu_{LL}$ as a function of moiré lattice filling $\nu$ follows a linear dependence (blue line). The moiré density is determined from the slope to be $n_M \approx (4.25 \pm 0.03) \times 10^{12}$ cm$^{-2}$.

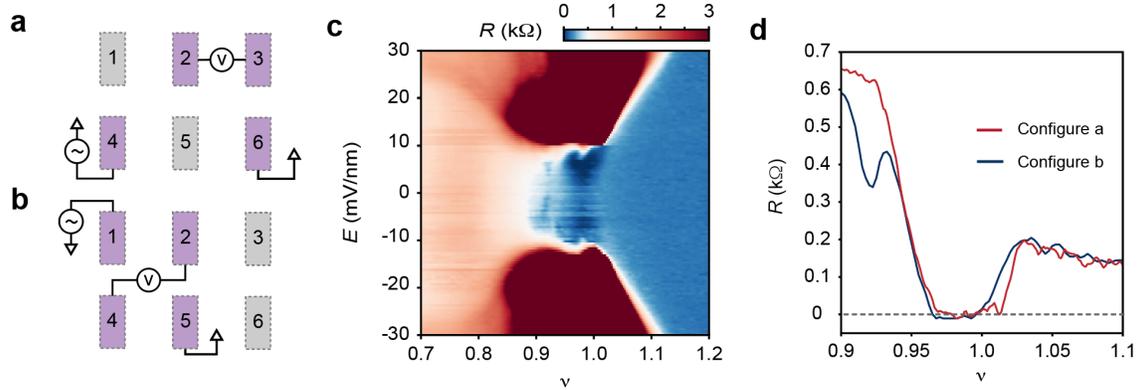

**Extended Data Figure 3 | Superconductivity in different measurement configurations. a,b,** Measurement configuration for four-terminal resistances: $R_{46,23}$ (**a**) and $R_{15,24}$ (**b**). A bias current is applied on the first two electrodes and the voltage drop is measured using the second two electrodes. The electrodes are labelled as in Extended Data Fig. 1b. Results in the main text are obtained using configuration **a**. **c,** Longitudinal resistance $R$ as a function of $E$ and $\nu$ at 50 mK and zero magnetic field using configuration **b**. **c,** Filling dependence of longitudinal resistance $R$ for measurement configuration a and b at $E \approx 8$ mV/nm and $T = 50$ mK. Independent of the measurement configuration, superconductivity is observed in tWSe$_2$ near $\nu = 1$.

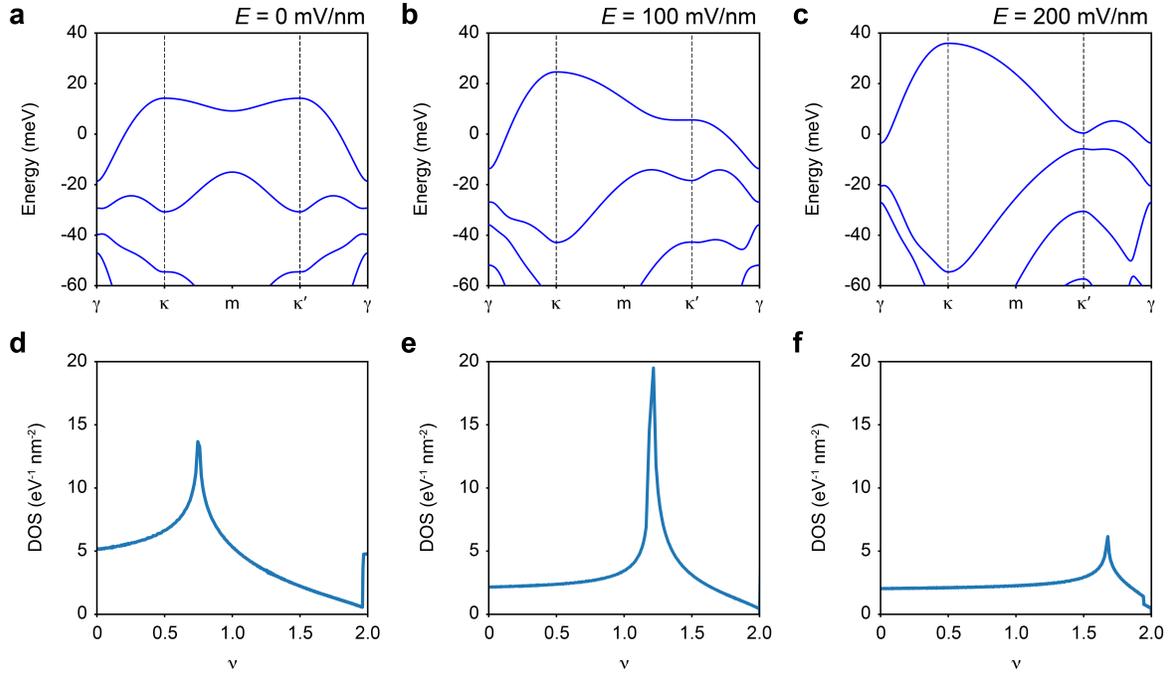

**Extended Data Figure 4 | Electronic band structure of 3.65°-tWSe$_2$ from the continuum model. a-c,** Topmost moiré valence bands of the K-valley state for $E = 0$ mV/nm (**a**), 100 mV/nm (**b**) and 200 mV/nm (**c**). **d-f,** The corresponding electronic DOS as a function of filling $\nu$ for the first moiré band. The vHS moves from $\nu < 1$ at $E = 0$ mV/nm to $\nu > 1$ at $E = 200$ mV/nm.

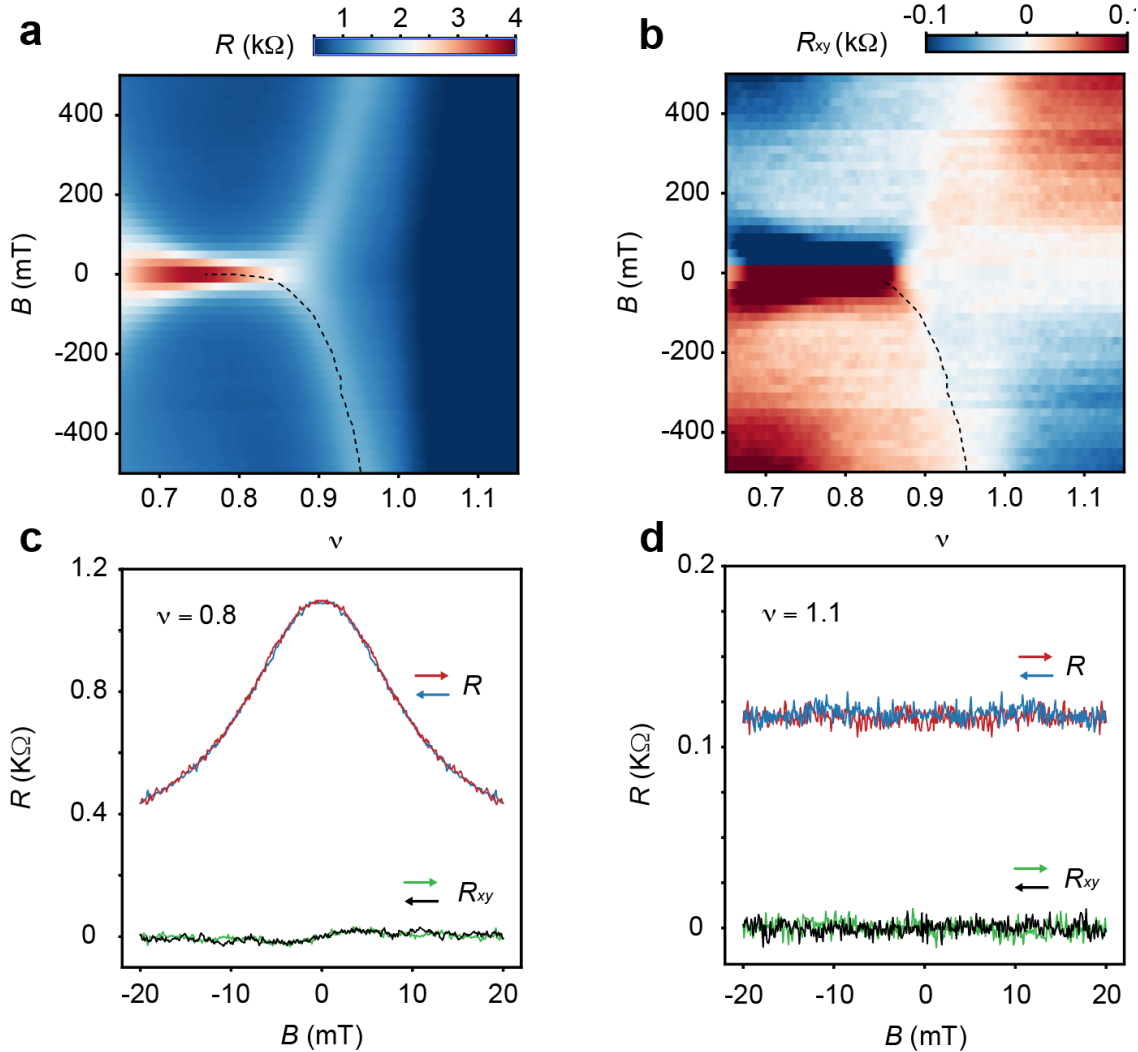

**Extended Data Figure 5 | Hall resistance. a,b,** Longitudinal resistance $R$ (**a**) and Hall resistance $R_{xy}$ (**b**) as a function of $B$ and $\nu$ at $T = 50$ mK and $E = 0$ mV/nm. Large bias (above the critical current) is applied, and superconductivity is not observed. The strong $R_{xy}$ response below filling 0.9 under small magnetic fields is an artifact because of the large magneto resistance and the coarse field step. The vHS manifests a peak in $R$ (**a**) and a sign change in $R_{xy}$ (**b**). The dashed lines are a guide to the eye of the location of the vHS for negative magnetic fields. The vHS is located at $\nu < 1$ for $B = 0$ and rapidly disperses with $B$ likely due to the combined Zeeman and orbital effects. **c,d,** Linecut of **a,b** at $\nu = 0.8$ (**c**) and $\nu = 1.1$ (**d**) with fine field scans. Both forward and backward field scans are displayed. Magnetic hysteresis is not observed. Anomalous Hall effect is also not observed ($R_{xy} = 0$). The artifact in **b** is removed under fine field scans.

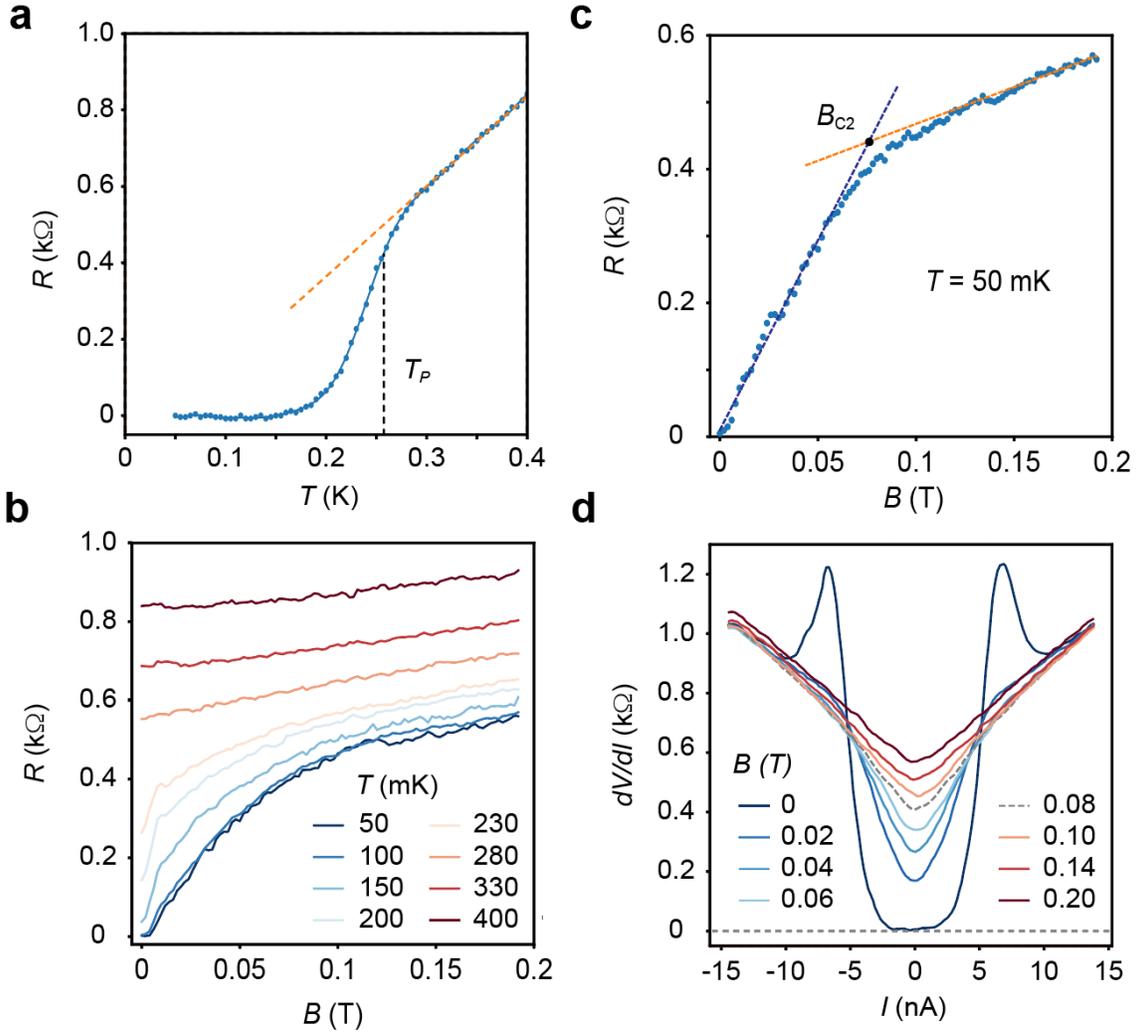

**Extended Data Figure 6 | Determination of $T_P$ and $B_{c2}$. a**, Temperature dependence of the zero-bias resistance $R$ at $\nu \approx 1$ ($E \approx 8$ mV/nm and $B = 0$). The pairing temperature $T_P$ ($\approx 250$ mK, vertical dashed line) is defined as the temperature, at which the measured resistance (blue line) deviates from the projected normal-state resistance (orange line) by 20%. Blue line: polynomial fit to the data (symbols); orange line: linear fit to the normal-state resistance ranging from 300 - 400 mK. **b**, Magnetic-field dependence of the longitudinal resistance $R$ at differing temperatures ($\nu \approx 1$ and $E \approx 8$ mV/nm). **c,** Magnetic-field dependence of $R$ at 50 mK. The critical field $B_{c2}$ ($\approx 80$ mT) is defined as the magnetic field, at which the orange and black dashed lines cross. Here the orange line is a linear fit to the normal-state resistance, and the black line is a fit of the unpinned vortex model, $R \propto \dfrac{B}{B_{C2}}$, to experiment for $B < B_{C2}$. **d,** Bias dependence of the differential resistance $\dfrac{dV}{dI}$ at varying magnetic fields ($\nu = 1$, $E \approx 8$ mV/nm and $T = 50$ mK). The dashed line corresponds to $B_{c2}$.

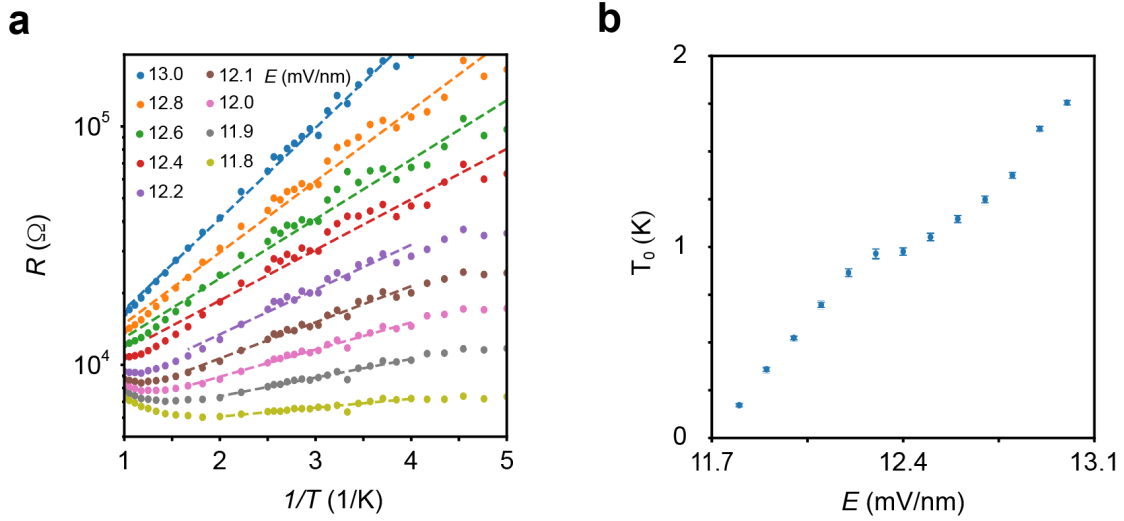

**Extended Data Figure 7 | Thermal activation analysis. a,** Arrhenius plot of the longitudinal resistance $R$ at varying $E$ (for $\nu = 1$ and $B = 0$). The dashed lines show the thermal activation fit and the range of data where the fit is good. **b,** Extracted gap size $T_0$ from **a** as a function of $E$ near $E_c \approx 11.7$ mV/nm. The gap vanishes continuously as $E$ approaches $E_c$ from above.